# Development of a TDC to equip a Liquid Xenon PET prototype

O. Bourrion, L. Gallin-Martel

*Abstract*—A Time to Digital Converter was designed (CMOS 0.35 μm) in order to be used in Liquid Xenon PET prototype. The circuit proved to be able to work at -120°C, while showing a resolution of 250 ps. The circuit enables a low readout dead time (<90 ns) and provides a fully synchronous digital interface for easy data retrieval.

## I. INTRODUCTION

Positron Emission Tomography (PET) can perform functional imaging of biological systems at molecular level. This technique consists in detecting 2 gamma photons emitted in opposite directions and in coincidence (<1 ns).

Since 2001 a R&D project has been initiated on the use of Liquid Xenon (LXe) for small animal tomography [1]. The advantage of this scintillator is to have a high light efficiency, and also, since it is liquid, the possibility to shape the detector at will. The shape of a module Fig. 1 is under study. On the contrary of crystal detector, modules can be distributed axially of the Field Of View.

When a gamma particle hits the Xenon, light is emitted on all direction and the photomultiplier tube or Avalanche Photodiode (also working at the temperature of LXe) converts it in an electrical signal. For allowing an image reconstruction, the module must provide the energy, localization and the date of the particle interaction; this means that a lot of information has to cross the cryostat. The solution to that problem is to develop an integrated front-end electronics close to the module at the LXe temperature and thus minimizing the number of wires that crosses this frontier (see Fig. 1).

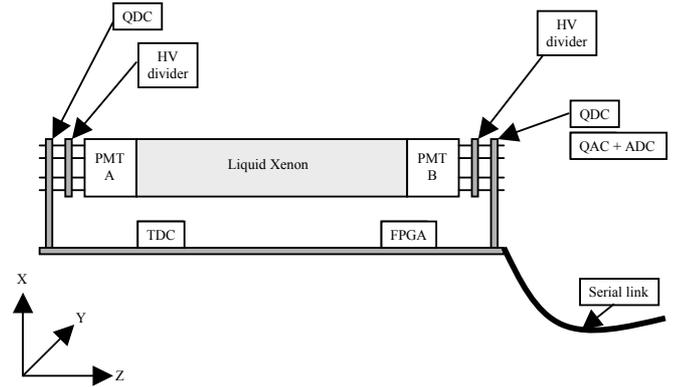

Fig. 1 Shape of a module and nearby electronics

The front end electronics comprises:
- 2 multi-channel charge to amplitude converters and 2 ADC (for the energy measurement)
- one Time to Digital Converter (for interaction dating)
- one FPGA for the serial link, ASIC general management, (x, y, z) coordinate calculation and module auto triggering

With this architecture, the only wires going out of the cryostat are the power supplies and the twisted pair serial link.

## II. DESIGN OF THE TDC

### A. Background

The aim of this project was not to design a TDC with a high resolution (see [2] for the different kinds of TDC), but a functional one for LXe. This implies that it must operate at 165 K, with resolution lower than 1 ns, and minimum current consumption (and therefore a self heating) kept to the minimum and a low dead time. All these criteria lead to the choice of a TDC based on a Delay Locked Loop, Fig. 2.

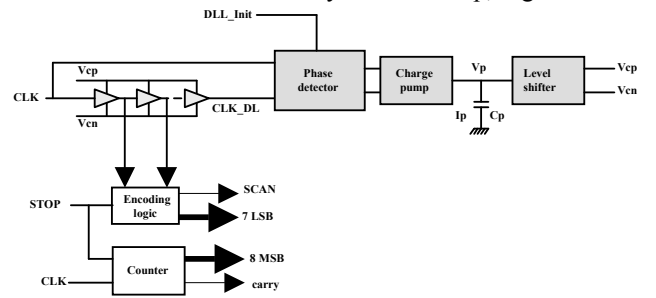

**Fig. 2 DLL overview**

The TDC uses the 128 elementary cells in the delay line in order to propagate the clock signal (here F=32MHz → T=31.25 ns). Once a hit pulse is received the state of the delay line is memorized and then encoded; after these steps, the position of the edge of the clock and the clock cycle number are provided on the digital interface, as well as a synchronous signal flagging the data.

In order to keep the same elementary delay over the process and temperature variations, the Delay Line has to be regulated, and thus becomes a Delay locked Loop.

### B. Delay Locked Loop

As shown in Fig. 3, the DLL is composed of 128 elementary delays, a phase detector, a charge pump and a level shifter.

*1) Elementary delay cell*

This cell comprises 2 inverters whose propagation time is voltage controlled by Vcp and Vcn.

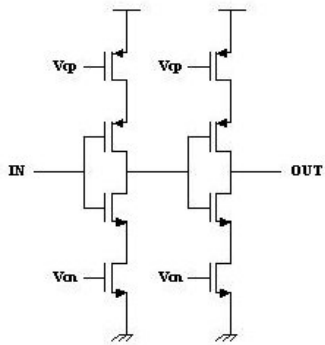

**Fig. 3 Voltage controlled gate**

A particular attention was paid to the clock duty cycle conservation [4], this is why the cell is highly symmetric.

*2) Phase detector*

The purpose of this block is to provide phase difference information for performing the line regulation.

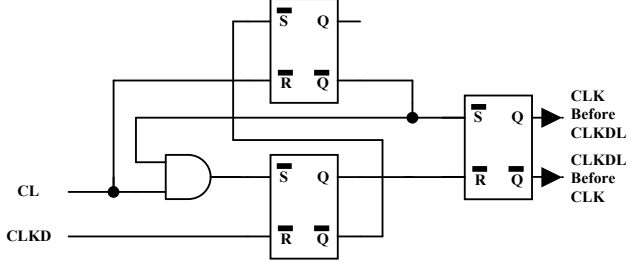

**Fig. 4 Phase detector**

With this kind of detector (Fig. 4), the delay between CLKDL and CLK is oscillating around the mean value of the delay: one clock period. This leads to the intrinsic jitter of the DLL (Fig. 5). Other kinds of phase detector are available but they exhibit other drawbacks [5].

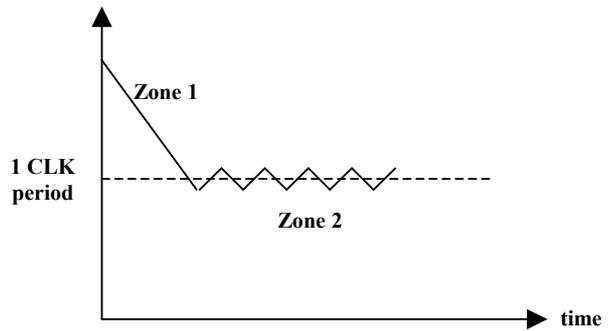

**Fig. 5 Line delay behaviour versus time**

During the implementation of this bloc, care was taken not to insert a phase offset with a non symmetrical laying out.

*3) Charge pump and shifter*

The charge pump is fed with the signals provided by the previous block for charging or discharging a capacitor (Fig. 6).

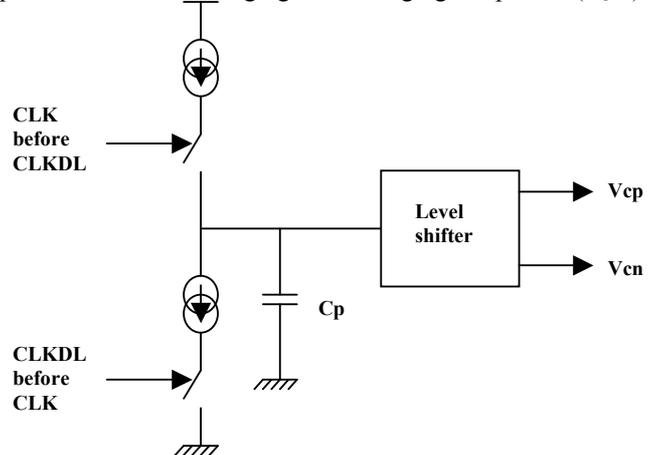

**Fig. 6 charge pump and level shifter**

The capacitor generates the control voltages through the shifter bloc. The shifter bloc is here for the linearization of the control signal because the function delay=f(Vcp,Vcn) is highly non linear.

### C. Line memorization

In order to have a proper line memorization and to avoid a false encoding, a buffer is inserted between the delay line and the memorization line. Therefore, a sufficient high slew rate signal is presented at the input of the latch and thus it decreases the probability to observe a metastable cell.

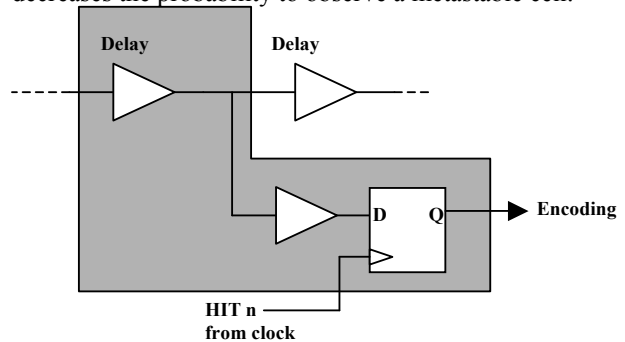

The clock tree distributed to the memorization cells has to be perfectly balanced.

*D. Digital part architecture*

At the input, the trigger signal is gated by a latch which does not allow a new trigger as long as the previous data has not been treated (Fig. 7).

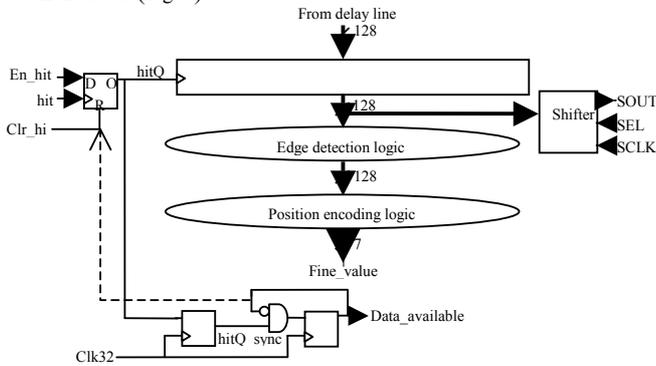

**Fig. 7 Digital architecture overview**

Also, the accepted trigger is resynchronized through two consecutive latches for generating a data available flag and thus providing an easy readout interface. The readout dead time is due to that (Fig. 8).

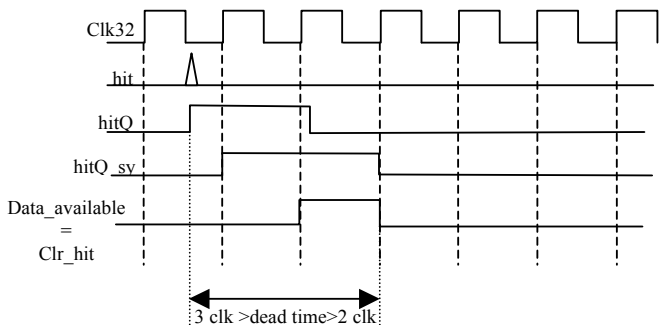

**Fig. 8 Data_available timing**

*1) Position encoding methodology*

For generating an encoded value of the fine delay, two steps are needed. At first, the clock rising edge must be localized in the memorized delay line, and then having this position (1 bit at one among 128 bit) the corresponding code must be provided.

If a close look is taken at the state of an eight elements delay line (Fig. 9), we can see that the position number is determined by the number of the last bit at 1 before the 1 to 0 change in the memorized line state. For instance when hit #1 occurs, the only delay element having its signal rising is Dly5, and the corresponding code (0111**1**000) yields bit 5.

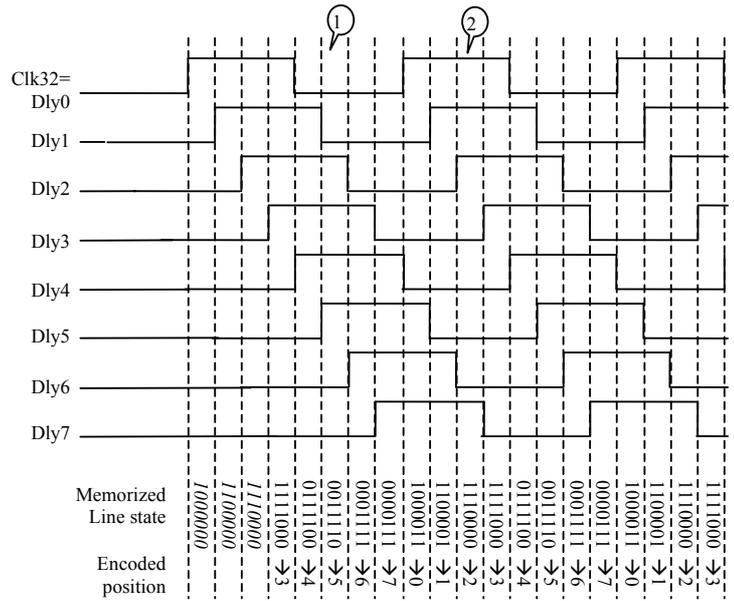

**Fig. 9 Eight elements delay line timing**

The digital system that allows detecting this edge in a line is given below Fig. 10 (replicated 128 times), it also provides a safety against defect in the memorized line due to metastability:

**Fig. 10 Edge detection logic**

Once the edge is located (only one bit at one in the line), the position has to be encoded. It can be easily done with OR gates. Below is given the methodology for an 8 bit line.

| Pos X | Bit 2 | Bit 1 | Bit 0 |
|---|---|---|---|
| 0 | 0 | 0 | 0 |
| 1 | 0 | 0 | 1 |
| 2 | 0 | 1 | 0 |
| 3 | 0 | 1 | 1 |
| 4 | 1 | 0 | 0 |
| 5 | 1 | 0 | 1 |
| 6 | 1 | 1 | 0 |
| 7 | 1 | 1 | 1 |

One can notice the following:
- $Bit_0$ = Pos1 **or** Pos3 **or** Pos5 **or** Pos7 ;
- $Bit_1$ = Pos2 **or** Pos3 **or** Pos6 **or** Pos7 ;
- $Bit_2$ = Pos4 **or** Pos5 **or** Pos6 **or** Pos7 ;

This can be easily extended for a 128 bit.

*2) Coarse counting*

Once the Hit has been encoded at a resolution of 250 ps, it must still be associated with the proper clock cycle. For doing that we use the architecture proposed by C. Lujslin [3] was used, where 2 counters are counting, one on the rising edge of the clock and the other one on the falling edge (Fig. 11).

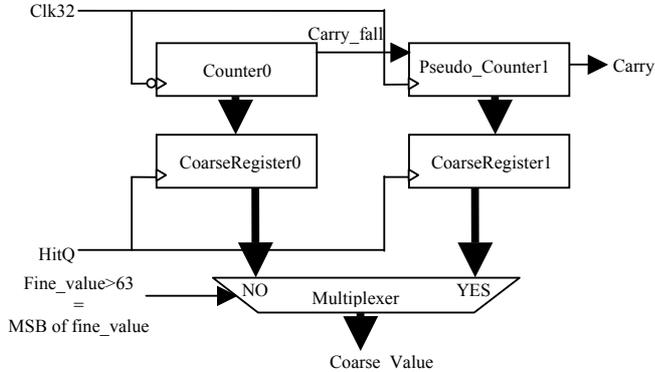

**Fig. 11 Coarse counting architecture**

Then, when a hit occurs, both counter values are saved, and the relevant one (stable at the sampling time) is multiplexed to the output. The multiplexer is controlled by the MSB of the fine value. A carry output is provided for counter range extension.

### III. CHIP TEST RESULTS

5 different circuits were tested.

- **Resolution :** 244 ps with a clock at 32MHz
- **DLL lock range :** from 20MHz to 40MHz
- **Differential non linearity:** +/- 20% (0.2 LSB) (excepted for channel 126 and 127, due to a layout problem).
- **Integral non linearity :** <1%
- **Memorized line state serial monitoring is functionnal**
- **Reproducibility :** For the same delay, the 5 ASIC give the same fine value
- **Consumption** : 14mW (190 µA on the 3V3 analog and 3,9mA on the 3V3 digital)
- **Pulse mid height width** : less than 2 channels

Testing in cryostat showed that the TDC could work a least to -120°C. The TDC parameter remained unchanged.

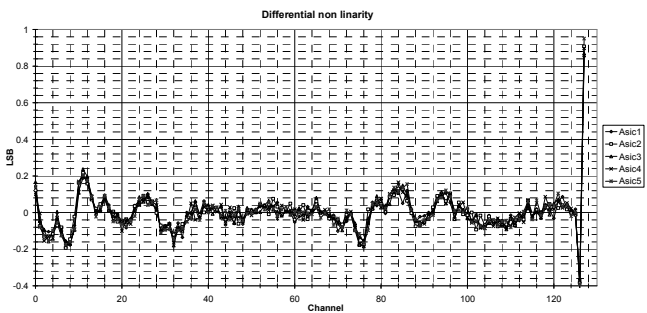

**Fig. 12 Differential non linearity plot for the 5 ASIC tested**